\documentstyle[12pt]{article}

\def\a{\begin{eqnarray}}
\def\b{\end{eqnarray}}
\def\0{\nonumber}
\input amssym.def
\font\teneusm=eusm10                    
\font\seveneusm=eusm7                   
\font\fiveeusm=eusm5                    
\newfam\eusmfam
\textfont\eusmfam=\teneusm
\scriptfont\eusmfam=\seveneusm
\scriptscriptfont\eusmfam=\fiveeusm

\def\sG{sine--Gordon}
\def\shG{sinh--Gordon}

\def\hRs{hyperelliptic Riemann surface}
\def\hRss{hyperelliptic Riemann surfaces}
\def\ags{algebraic--geometrical solution}
\def\agss{algebraic--geometrical solutions}
\def\dg{ dressing group}
\def\dge{ dressing group element}

\def\dgt{ dressing group transformation }
\def\dt{ dressing transformation}
\def\dts{ dressing transformations}
\def\PLa {Poisson--Lie action }

\renewcommand{\theequation}{\thesection.\arabic{equation}}

\newlength{\extraspace}
\newlength{\extraspaces}
\newcounter{dummy}

\newcommand{\ai}{
\addtocounter{equation}{1}
\setcounter{dummy}{\value{equation}}
\setcounter{equation}{0}
\renewcommand{\theequation}{\thesection.\arabic{dummy}\alph{equation}}
\begin{eqnarray}
\addtolength{\abovedisplayskip}{\extraspaces}
\addtolength{\belowdisplayskip}{\extraspaces}
\addtolength{\abovedisplayshortskip}{\extraspace}
\addtolength{\belowdisplayshortskip}{\extraspace}}
\newcommand{\bj}{
\end{eqnarray}
\setcounter{equation}{\value{dummy}}
\renewcommand{\theequation}{\thesection.\arabic{equation}}}

\input amssym
\newcommand{\be}{\begin{equation}}
\newcommand{\ee}{\end{equation}}
\newcommand{\ba}{\begin{eqnarray}}
\newcommand{\ea}{\end{eqnarray}}
\newcommand{\ban}{\begin{eqnarray*}}
\newcommand{\ean}{\end{eqnarray*}}
\newcommand{\brr}{\begin{array}}
\newcommand{\err}{\end{array}}
\newcommand{\bc}{\begin{center}}
\newcommand{\ec}{\end{center}}

\def\A{{\cal A}}
\def\B{{\cal B}}
\def\C{{\cal C}}
\def\D{{\cal D}}
\def\E{{\cal E}}

\def\H{{\cal H}}
\def\N{{\cal N}}

\def\L{ \Lambda}
\def\N{{\cal N}}

\def\O{{\cal O}}

\def\T{{\cal T}}

\def\l{\lambda}

\def\al{\alpha}
\def\be{\beta}
\def\de{\delta}
\def\ep{\epsilon}

\def\var{\varphi}
\newcommand{\bea}{\begin{eqnarray}}
\newcommand{\eea}{\end{eqnarray}}
\newcommand{\bean}{\begin{eqnarray*}}
\newcommand{\eean}{\end{eqnarray*}}
\newcommand{\del}{\partial}

\newcommand{\dpiu}{\partial_{x^+}}
\newcommand{\dmeno}{\partial_{x^-}}

\begin{document}

\begin{titlepage}

\begin{flushright}
SISSA-ISAS 176/94/EP

\end{flushright}
\vskip0.5cm
\centerline{\LARGE   Dressing  Transformations and the }
\centerline{\LARGE Algebraic--Geometrical Solutions }
\centerline{\LARGE in the Conformal Affine $sl(2)$ Toda Model}
\vskip1.5cm
\centerline{\large   R. Paunov}
\centerline{International School for Advanced Studies (SISSA/ISAS)}
\centerline{Via Beirut 2, 34014 Trieste, Italy}
\centerline{and INFN, Sezione di Trieste.  }
\vskip5cm
\abstract{ It is  shown that the  algebraic--geometrical (or quasiperiodic)
 solutions of the Conformal
Affine $sl(2)$ Toda model are generated from the vacuum via dressing
transformations. This result generalizes the result of Babelon and Bernard
which states that the $N$--soliton solutions are generated from the vacuum
by dressing transformations.}

\end{titlepage}

\section{Introduction}

\setcounter{equation}{0}
\setcounter{footnote}{0}

The \dg\,  \cite{Sem}, \cite{BB} is a  symmetry of the nonlinear equations
which admit a zero--curvature representation. It acts via  gauge
transformations which leave invariant the form of the Lax connection and
hence the dressing symmetry acts on the space of solutions of the
corresponding integrable model.
An important property of the \dg\, is that its action is a  \PLa .
This means that in order to ensure the covariance of the Poisson brackets
one has to introduce a nontrivial bracket on the \dg.
It was argued in \cite{BB} that the \dg\,
appears as  a semiclassical limit of the quantum group symmetry of the
two--dimensional integrable quantum field theories.

In this paper we study the relation between the \dts\, and the  \agss\,
(or finite--gap solutions)
of the Conformal Affine Toda (CAT) $sl(2)$ model  which is a
conformally invariant extension of the \shG\, equation \cite{BaBo}.
 The \agss\, of the
\sG\, equation ( which differs from the \shG\,  by a trivial renormalization)
 are studied extensively in the literature \cite{Mam}-\cite{Harnad}.
The $N$--gap \shG\, solutions are related  to \hRss\, of genus $2N-1$
with a fixed--point--free automorphism.
This relation can be expressed explicitly
by the evolution of the $N$--gap \shG\, system (\ref{Jacobi}),(\ref{epphi})
or equivalently by the flow equations (\ref{moto}).
In the singular limit when the length
of all branch cuts of the underlying \hRs\, tend to zero, the $N$--gap
solutions coincide with the $N$--soliton solutions \cite{Date},
\cite{Fad}, \cite{Nov}. The solitons, as
it was shown by Babelon and Bernard \cite{BaBe},  belong to the \dg\,
orbit of the vacuum.  The \dg\, elements
which generate the soliton solutions from the vacuum are
found.  The $N$--soliton solutions of the  \sG\, model are considered
 as a relativistically invariant $N$--body problem in \cite{NBB}.

We outline the content of the paper. Sec. 2 is a review of the dressing
symmetries of the $sl(2)$ CAT  and \shG\, models. The sec. 3 deals with the
\agss\, of the \shG\, equation and the related to them dressing problem.
The \dg\, elements which transform the vacuum into the $N$--gap solutions
are constructed starting from the explicit expression for the transport
matrix (\ref{trans}).
Sec. 4 is devoted to the
dressing problem  for the \agss\, of the $sl(2)$ CAT model.

\section{Dressing transformations in the $sl(2)$ CAT and the \shG\, models}

\setcounter{equation}{0}
\setcounter{footnote}{0}

Both the $sl(2)$ CAT and the \shG\, models appear as a zero--curvature
condition of Lax connections the components of which belong to the  affine
$\hat{sl}(2)$ algebra and the $\tilde {sl}(2)$ loop algebra respectively.
We  introduce some Lie algebra notations.
Denote by $E^{\pm}$ and $H$ the three generators of the $sl(2)$ Lie
algebra with the commutators
\a
[ H , E^{\pm} ]= \pm 2 E^{\pm}~~~~~~~[E^+ , E^-]=H
\b
In the fundamental (two--dimensional) representation we shall fix these
generators by normalizing $H=$ diag$(1, -1)$. The
loop algebra $\tilde{sl}(2)$ is the Lie algebra of the traceless $2 \times
2$ matrices whose entries are Laurent series in the spectral parameter
$\l$; moreover, in the principal gradation, the diagonal elements contain
only even powers of $\l$ while the off--diagonal entries contain only odd
powers  of the spectral parameter.
The loop algebra ${\tilde sl}(2)$ has a central
extension which is the affine $\hat{sl}(2)$ algebra. As a basis
of $\hat{sl}(2)$ in the principal gradation
one can choose the elements: $H_n=\l^n H$ where $n$ is an even integer,
$E^{\pm}_n=\l^n E^{\pm}$ with odd integer $n$, the derivation $\hat d$ and
the central element $\hat c$. The commutation relations are
\a
[ H_m , E^{\pm}_n ]=\pm 2E^{\pm}_{m+n}~~~~~~~~~
[ H_m , H_n ] = m \hat{c} \de_{m+n} \0
\b
\a
[ E^{+}_m , E^{-}_n ] = H_{m+n} + \frac{\hat{c}}{2}m \de_{m+n}\0
\b
\a
[ \hat{d} , X_n ] = n X_n ~~~~~~~~~ X_n=\l^n X, ~~~~~~ X=H, E^{\pm}
\label{affine}
\b
and $\hat{c}$ commutes with all the generators. The derivation $\hat{d}$
defines a gradation in $\hat{sl}(2)$: the element $X_n=\l^n X,\,\,\,
X\in sl(2)$ has  grade $n$. We also recall the Cartan decomposition:
$\hat{sl}(2)=\N_+\oplus \N_- \oplus \H $ where $\N_+$ ( $\N_-$ )
is spanned on  the positive (negative) grade elements  and $E_0^{+}$
( $E_0^{-}$ ); the Cartan subalgebra $\H$ has three independent elements:
$H$, $\hat{d}$ and $\hat{c}$.
The Borel subalgebras are $\B_{\pm}=\H \oplus \N_{\pm}$.
The highest weight vectors $|\L>$ are
annihilated by $\N_+$ and are eigenvectors of the elements of the Cartan
subalgebra.  In what follows
we shall frequently use the notation $\E_{\pm}= \l^{\pm 1}
(E^++E^-)$.

The equations of motion of the $sl(2)$ CAT model follow from the flatness
of the connection
\ai
\D_{x^{\pm}}&=& \del_{x^{\pm}}+A_{x^{\pm}}=
\del_{x^{\pm}} \pm \del_{x^{\pm}}\Phi + m e^{\pm ad \Phi} \E_{\pm}~~~~~~~
\del_{x^{\pm}}=\frac{\del}{\del x^{\pm}}
\label{con}\\
\Phi&=& \frac{1}{2} \var H + \eta \hat{d} + \frac{1}{4} \zeta \hat{c}
\label{phi}
\bj
The zero--curvature representation of the \shG\, equation has the same
form as (\ref{con}) with $A_{x^{\pm}}$ being in the loop algebra
$\tilde{sl}(2)$ and $\Phi=\frac{1}{2} \var H $. Inserting (\ref{con}) and
(\ref{phi}) in $[\D_{x^+} , \D_{x^-}]=0$ one obtains the $sl(2)$
CAT equations \cite{BaBo}
\ai
\dpiu\dmeno \var&=&m^2e^{2\eta}\left( e^{2\var}- e^{-2\var} \right)
\label{sinh}\\
\dpiu\dmeno \eta &=& 0 \label{eta}\\
\dpiu\dmeno \zeta &=& m^2 e^{2\eta}\left( e^{2\var}+ e^{-2\var} \right)
\label{zeta}
\bj
the first of which reduces to the \shG\, equation for $\eta=0$; $m$
is the mass of the \shG\, field.  As in \cite{BaBe}
we shall deal with solutions of (\ref{sinh})--(\ref{zeta}) with
$\eta=0$.

The \dt s are gauge transformations which  preserve the form of the
connection (\ref{con}).
Since $A_{\pm} \in \B_{\pm}$ one  requires that the dressing transformation
$\Phi \rightarrow \Phi^g$  is induced by a couple of gauge transformations
$g_{\pm} \in e^{\B_{\pm}}$
\a
A^g_{\mu}= -\del_{\mu} g_{\pm} g_{\pm}^{-1}+ g_{\pm} A_{\mu} g_{\pm}^{-1}
{}~~~~~~~ \mu=x^+, x^-
\label{gauge}
\b
where  $A^g_{\pm}$ are given by (\ref{con}) but with $\Phi$ replaced by
$\Phi^g$. Let $T(x^+, x^-,\l)$ be the normalized transport matrix
associated to
(\ref{con}), i. e. $\D_{x^{\pm}} T=0$ and $T(0,0,\l)=1$. The \dg\, elements
$g_{\pm}$ are then solution of the factorization problem
\a
g_-^{-1}(x^+, x^-,\l) g_+ (x^+, x^-, \l )= T(x^+, x^-,\l)
g_-^{-1}(0, 0,\l) g_+ (0, 0, \l ) T^{-1}(x^+, x^-,\l)
\label{fat}
\b
which reflects the requirement that the transformation $T \rightarrow T^g$
can be performed by using either $g_+$ or $g_-$. It is also not difficult
to see that $g_+$ and $g_-$ have inverse components on the
Cartan subgroup \footnote{more precisely the elements $g_{\mp}$ in the
affine group are fixed up to the factors $exp \{ f_{\pm}(x^{\pm}) \hat{c}\}$
where $f_{\pm}$ are arbitrary functions}
\a
g_{\mp}=e^{\pm(\Phi^g-\Phi)} \cdot e^{X_{\mp 1}} \cdot e^{X_{\mp 2}} \ldots
e^{X_{\mp i}} \ldots
\label{prod}
\b
where the grade of the elements $X_{\pm i}$ is $\pm i$. Inserting the
upper expansion in (\ref{gauge}) one obtains infinite set of equations
which characterize the \dg\, orbit of the solution $\Phi$. The first few
equations are
\ai
&&m [ X_{\mp 1 }, e^{\pm ad \Phi} \E_{\pm} ] = \pm 2 \del_{x^{\pm}}
\left( \Phi^g- \Phi \right) \label{uno}\\
&&m [ X_{\mp 2 }, e^{\pm ad \Phi} \E_{\pm} ] = \del_{x^{\pm}} X_{\mp 1}
\mp [ X_{\mp 1} , \del_{x^{\pm}} \Phi^g ] \label{due}\\
&&\del_{x^{\pm}} X_{\pm 1}\mp [ X_{\pm 1} , \del_{x^{\pm}} \Phi ] =
m\left( e^{\pm ad \Phi} - e^{\pm (2\Phi^g-\Phi)} \right) \E_{\pm}\label{tre}
\bj
which are valid both for the $sl(2)$ CAT and \shG\, models
since in the derivation  one uses only the grade analysis. Note that
(\ref{uno}) fixes  $X_{\pm 1}$ uniquely in the affine algebra
\a
X_{\pm 1}&=& \frac{e^{-\var}}{2m}
\left( \del_{x^{\mp}} \var + \del_{x^{\mp}}\zeta - \del_{x^{\mp}} \var^g -
\del_{x^{\mp}} \zeta^g \right) \l^{\pm 1} E^{\pm}+\0\\
&&+\frac{e^{\var}}{2m}\left( - \del_{x^{\mp}} \var + \del_{x^{\mp}}\zeta +
\del_{x^{\mp}} \var^g-\del_{x^{\mp}} \zeta^g\right) \l^{\pm 1} E^{\mp}
\label{Xuno}
\b
while in the loop algebra due to the absence of a central element
(\ref{uno}) leaves a one--parameter freedom in $X_{\pm 1}$. Further,
due to the fact that the unique independent element
of grade $\pm 2$ is $\l^{\pm 2}H$,
the equation (\ref{due}) implies a new restriction on
$X_{\pm 1}$ which together with (\ref{Xuno}) gives
\a
\left( \del_{x^{\pm}} \var^g \right)^2 - \del_{x^{\pm}}^2\zeta^g=
\left( \del_{x^{\pm}} \var \right)^2 - \del_{x^{\pm}}^2\zeta
\label{inv}
\b
and hence the quantity $(\del_{x^{\pm}} \var)^2-\del_{x^{\pm}}^2\zeta$ is an
invariant of  the dressing group orbit. An important observation
done by Babelon and Bernard in \cite{BaBe} is that
the central element $\hat{c}$ appears only in
(\ref{uno}) ( and yields the solution (\ref{Xuno})) and all the other
restrictions on $X_{\pm i}$ can be calculated in the loop algebra. Therefore,
if $\tilde{g}_{\pm}$ is a solution of the dressing problem (\ref{gauge})
for the \shG\, solutions $\var$ and $\var^g$, then the elements of the affine
$\hat{SL}(2)$ group
\a
\hat{g}_{\pm} (x^+, x^-,\l )= e^{\mp \frac{\zeta^g-\zeta}{4}\hat{c}}
\tilde{g}_{\pm} (x^+, x^-,\l )
\label{trucco}
\b
define a \dt\, which relates the solutions $(\var , \zeta)$ and $(\var^g,
\zeta^g)$ of the $sl(2)$ CAT model. Note that $\zeta$ and $\zeta^g$ are
determined from (\ref{Xuno}) with $X_{\pm 1}$ derived from the grade
expansions (\ref{prod}) of the {\it loop group} elements
$\tilde{g}_{\pm}$. We shall use this remark in the demonstration
that the  \agss\, of the $sl(2)$ CAT model are in the \dg\, orbit of the
vacuum.

The  \dg\, action on the fields $\xi_{\L} =<\L|e^{\Phi},\,\,\,
\bar{\xi}=T^{-1}e^{-\Phi}|\L>$ where $\L$ is a highest weight is
extremely simple
\a
\xi^g_{\L}(x^+, x^-, \l)=\xi(x^+, x^-, \l)\cdot g_-^{-1}(0,0,\l)~~~~~~~~~
\bar{\xi}^g_{\L}(x^+, x^-, \l)=g_+ (0,0,\l) \cdot \bar{\xi}_{\L}(x^+, x^-,
\l)\0
\b
The last equation gives a nice relation between the tau--functions
$\tau_{\L}(\Phi)=e^{-2\L (\Phi)}$
of the solutions $\Phi$ and $\Phi^g$ \cite{Kyoto}
\a
e^{-2\L (\Phi^g(x^+,x^-))}&=&\xi_{\L}(x^+, x^-, \l)\cdot g \cdot
\bar{\xi}_{\L}(x^+, x^-, \l)\0\\
e^{-2\L (\Phi(x^+,x^-))}&=&\xi_{\L}(x^+, x^-, \l)\cdot
\bar{\xi}_{\L}(x^+, x^-, \l)\0\\
g&=&g_-^{-1}(0,0,\l) \cdot g_+ (0,0,\l)
\label{tau}
\b
The above relations show that the multiplication in the \dg\, is different
from the multiplication in the initial loop or affine group: the product
of two elements $g=g_-^{-1} g^+$ and $h=h_-^{-1} h_+$ is
$g \times h=(g_- h_-  )^{-1} \cdot g_+ h_+$. We note also that the
correspondence $g \rightarrow (g_-, g_+)$
is one--to--one since due to (\ref{prod}) $g_-$ and $g_+$ have inverse
components on the Cartan subgroup.

\section{ The dressing problem for the \agss\, of the \shG\, equation}

\setcounter{equation}{0}
\setcounter{footnote}{0}

The algebraic--geometrical solutions of the \shG\, equation are
expressed in terms of theta functions on \hRss\, of odd genus $2N-1$
which possess a fixed--point--free automorphism of order two \cite{Date}.
 There is another expression  \cite{Mam}, \cite{DN}, \cite{Harnad}
based on theta functions on arbitrary \hRss .
In \cite{P} it is demonstrated that these two expressions are in fact
equivalent. In this paper we shall use another description:
in the spirit of \cite{NBB}, the
\agss\, of the \shG\, equation will be represented  as solutions of
an integrable system with finite number of degrees of freedom. From  this
point of view the \hRss\, appear
as characteristic equations of the corresponding Lax operators which turn
out to be  polynomials of finite order on the spectral parameter
\cite{Harnad}, \cite{Kr}.
Let $\hat{\C}_N$ be the \hRs\, of the algebraic function $s_N=s_N(\mu)$
given by the equation
\a
s_N^2= \prod_{p=1}^{2N}\left(\l^2-\mu_p^2 \right)
\label{sp}
\b
where $ \pm \mu_p \,\,\, p=1\ldots 2N$ are the ramification points.
The genus of $\hat{C}_N$ is $2N-1$. We note that $\hat{\C}_N$ has a free
involution
$T:(\l , s) \rightarrow (-\l , -s)$.  The \agss\, of the \shG\,
corresponding to (\ref{sp}) are introduced in terms of a positive
divisor of degree $N$ on $\hat{\C}_N$ $D(x)=\sum_{j=1}^N P_j(x)$
where $x=(x^+,x^-)$ and $P_j(x)=(\ep_j(x) , s(\ep_j(x))$. The evolution
of $D(x)$ is determined by
\ai
\sum_{j=1}^N
\int_{P_j(0)}^{P_j(x)} \frac{\mu^{2l} d\mu}{s_N(\mu)}&=&
\frac{m}{\prod_{p=1}^{2N} \mu_p} x^- \de_{l,0} +
(-)^{N-1}mx^+\de_{l,N-1} \0\\
l&=&0,1\ldots N-1 \label{Jacobi}\\
e^{\var_N}&=&\frac{\sqrt{\prod_{p=1}^{2N}\mu_p}}{\prod_{j=1}^N \ep_j}
\label{epphi}
\bj
{}From (\ref{Jacobi}) it follows that the $x^{\pm}$ flows of the system are
linear on the Jacobian variety of the \hRs\,  $\C_N$ of genus $N$  which is
the quotient of $\hat{\C}_N$ by  the  free automorphism T \cite{Harnad}.
{}From  (\ref{Jacobi}) one gets the system
\a
\dpiu \ep_j= ms_N(\ep_j)\prod_{l\neq j} \frac{1}{\ep_l^2-\ep_j^2}~~~~~~~~
\dmeno \ep_j= \frac{m}{\prod_{p=1}^{2N}\mu_p} s_N(\ep_j)
\prod_{l\neq j} \frac{\ep_l^2}{\ep_l^2-\ep_j^2}
\label{moto}
\b
which is integrable for arbitrary function $s_N$. The \shG\, equation
$\dpiu\dmeno \var_N=2m^2sh2\var_N$ is equivalent to the identity
\a
&&\sum_{j=1}^N\left( \frac{1}{\ep_j^2}- \sum_{p=1}^{2N}
\frac{1}{\ep_j^2-\mu_p^2} - \sum_{k\neq j } \frac{1}{\ep_k^2-\ep_j^2}\right)
s_N^2(\ep_j)\prod_{l\neq j} \frac{ \ep_l^2}{(\ep_l^2-\ep_j^2)^2}=\0\\
&&=\frac{\prod_{p=1}^{2N}\mu_p^2}{\prod_{j=1}^N \ep_j^2}
- \prod_{j=1}^N \ep_j^2
\label{iduno}
\b
We sketch its proof: first we notice that  both sides of (\ref{iduno})
are symmetric meromorphic functions on the variables $\ep^2_1\ldots \ep^2_N$;
the left hand side has no poles at the points $\ep^2_l=\ep^2_j$
( $l\neq j$ )  and   $\ep_l^2=\mu_p^2$. Due to the symmetry one can consider
the two sides of (\ref{iduno}) as  functions of, say $\ep_1^2$. It is easy
to see that their residues  at $\ep_1^2=0, \infty$ coincide
and therefore, their difference is a constant on $\ep_j^2$'s. In order to
show that this constant is zero we set $x_j=\mu_j^2$,
 and $y_j=\mu^2_{j+N}$ ($j=1\ldots N$) in  the identity
\a
\sum_{j=1}^N \prod_{k=1}^N (x_j-y_k)
\prod_{l\neq j}\frac{x_l}{(x_l-x_j)^2}=x_1\ldots x_N-y_1\ldots y_N
\label{iddue}
\b
In the singular limit  $\mu_{2j-1}\rightarrow\mu_{2j}=\al_j,\,\,\,
j=1\ldots N$
in (\ref{sp}) the integrals (\ref{Jacobi}) are expressed by
logarithms and the corresponding \shG\,   solution (\ref{epphi})
is  identical to the $N$--soliton solution. The parameters $\al_1 \ldots
\al_N$ are the soliton  rapidities. The soliton limit of the finite--gap
solutions will be discussed  in details in \cite{Pa}.

The hyperelliptic curve (\ref{sp}) is a two--sheeted covering of the
Riemann sphere by the spectral parameter $\l$ with ramification points
at $\pm \mu_1 \ldots \pm \mu_{2N}$. We shall assume in what follows that
the ramification points are all finite and different from zero
\footnote{In order to get a non--singular \hRs\, one should also assume
that $\mu_i\neq \mu_j$ for $i\neq j$ and $\mu_i+\mu_j \neq0$}.
 Then
$\hat{C}_N$ is represented as a couple of Riemann spheres $\hat{C}^+_N$
and $\hat{C}^-_N$ glued along $2N$ cuts connecting the branching points.
For example, as cuts one can choose the lines between $\pm \mu_{2j-1}$
to $\pm \mu_{2j}$. Denote the infinity and the zero points on
$\hat{C}^{\pm}_N$ as $\infty_{\pm}$ and $\O_{\pm}$ respectively.
Directly from
(\ref{sp}) one gets the expansions
\ai
s_N&=&\pm \l^{2N} \left( 1 + O(\frac{1}{\l^2}) \right) ~~~~~~~
(\l,s) \rightarrow \infty_{\pm}\label{inf}\\
s_N&=& \pm (-)^N \prod_{p=1}^{2N} \mu_p \left( 1 + O(\l^2)\right) ~~~~~~~
(\l, s ) \rightarrow \O_{\pm} \label{zero}
\bj
We continue with the problem of construction of a dressing transformation
which generates the solution $\var_N$ from the vacuum $\var_0=0$. In
\cite{Pa} it will be shown that the matrix
\a
&&\T_N(x,\l,s) =e^{\frac{\var_N}{2}H}\cdot\0\\
\hskip -1.5cm &&\left (\begin{array}{cc}
\prod_{j=1}^N \frac{\l+(-)^N\ep_j(x)}{\l+(-)^N\ep_j(0)} e^{\A_N}&
\prod_{j=1}^N \frac{\l+(-)^{N+1}\ep_j(x)}{\l+(-)^{N+1}\ep_j(0)} e^{-\A_N}\\
{}~&~\\
\frac{s_N(\l)-l_N(\l)}{\prod_{j=1}^N
(\l+(-)^{N+1}\ep_j(x))(\l+(-)^N\ep_j(0))}e^{\A_N}&
-\frac{s_N(\l)+l_N(\l)}{\prod_{j=1}^N
(\l+(-)^N\ep_j(x))(\l+(-)^{N+1}\ep_j(0))}e^{-\A_N}\end{array}
\right )\label{trans}
\b
where
\ai
\A_N&=&(-)^{N-1} s_N(\l) \left( \frac{mx^-}{\l \prod_{p=1}^{2N}\mu_p}+
\l\sum_{j=1}^N\int_{P_j(0)}^{P_j(x)}\frac{d
\mu}{(\mu^2-\l^2)s(\mu)}\right)+\0\\
&&+\frac{(-)^N}{2} ln \prod_{j=1}^N \frac{(\l-\ep_j(x))(\l+\ep_j(0))}
{(\l+\ep_j(x))(\l-\ep_j(0))}\label{Acal}\\
l_N(\l)&=& (-)^N \l \sum_{j=1}^N \frac{s(\ep_j(x))}{\ep_j(x)}
\prod_{l\neq j} \frac{\l^2-\ep_l^2(x)}{\ep_j^2(x)-\ep_l^2(x)}
\label{ele}
\bj
is a  non--normalized transport matrix corresponding to the $N$ gap solution
(\ref{epphi}). Its determinant
\a
det \T_N = -2 \frac{ s_N(\l)}{\prod_{j=1}^N (\l^2-\ep_j^2(0))}
\label{det}
\b
vanishes at the ramification points.  In analysing the singularities of
(\ref{trans}) we first note that due to (\ref{inf}), the
integrated $x^+$ and $x^-$ flows of
the system (\ref{Jacobi}) are equivalent to the expansions $\A_N= \mp m\l x^+
+O(\frac{1}{\l})$  around the infinity points $\infty_{\pm}$ while from
(\ref{zero}) one immediately obtains
$\A_N= \mp \frac{mx^-}{\l}+O(\l)$ at the vicinity of the zero points
$\O_{\pm}$.  The integrand of the second term in (\ref{Acal})
is a  differential of a third kind with simple poles at the
points $(\l, \pm s)$ and $(-\l , \pm s)$. Therefore, one expects  $\A_N$
to have logarithmic  singularities at the points whose projections on the
sphere
by $\l$ are $\pm \ep_j(x)$ or $\pm \ep_j(0)$. Note that if such a
singularity exists at some of these points belonging to
$\hat{C}_N^+$ ($\hat{C}_N^-$) then at the
point on $\hat{C}_N^-$($\hat{C}_N^+$)  with the same value of the
spectral parameter $\A_N$ is {\it regular}.
Taking into account these remarks it is easy to see that
the elements of (\ref{trans}) are {\it functions} on $\hat{C}_N$ with simple
poles which does not depend on $x^+$ and $x^-$ and exponential singularities
 at the infinity and the zero points. Fixing all the points
$P_j(0),\,\,\, j=1\ldots N$ of the divisor $D(0)$ to belong to
$\hat{C}_N^+$ it is easy to see that on $\hat{C}_N^+$
the elements of  the first (the second) column of $\T_N$ have simple poles
at the points $\l=(-)^{N+1}\ep_j(0)$ ($\l=(-)^{N}\ep_j(0)$); in order to
find the positions of the poles on $\hat{C}_N^-$ it is sufficient to
note that the change $s\rightarrow -s$ in (\ref{sp}) exchanges the columns
of the transport matrix. We have thus shown that the entries of
(\ref{trans}) have the following property : if at certain point of
$\hat{C}_N$ which is not a ramification point there is a simple pole of a given
matrix element, then at the point with the same value of the spectral
parameter but with the opposite sign of $s$ the same matrix element has no
singularity. Similar property was established for the  Bloch functions
associated with the finite--gap   solutions of the KdV equation
which are also known to correspond to \hRss\, \cite{Nov}. In \cite{Date}
the transport matrix (\ref{trans}) was expressed in terms of theta
functions on $\hat{C}_N$.

The explicit knowledge of the  transport matrix $\T_N$ allows
to find a
 dressing transformation which generates the solution
(\ref{epphi}) from the vacuum. In doing that we express the normalized
transport matrix $T_N$ corresponding to the solution $\var_N$ as
\ai
T_N(x,\l, s)&=& \T_N(x,\l, s)  \T^{-1}_N(0,\l, s)=
f_N(x,\l, s) T_0(x,\l) f^{-1}_N(0,\l, s)\label{numero}\\
f_N(x,\l, s)&=&\T_N(x,\l, s)\T_0^{-1}(x,\l) S(\l , s)\label{fr}\\
\T_0(x,\l)&=&
\left (\begin{array}{cc}
e^{-m(\l x^++\frac{x^-}{\l})}& e^{+m(\l x^++\frac{x^-}{\l})}\\
e^{-m(\l x^++\frac{x^-}{\l})}& -e^{+m(\l x^++\frac{x^-}{\l})}
\end{array}
\right )\0\\
S(\l , s)&=&\left (\begin{array}{cc} a(\l ,s)&b(\l ,s)\\
b(\l , s)&a(\l ,s)\end{array}
\right )~~~~~~ a^2-b^2=\frac{\prod_{j=1}^N(\l^2-\ep_j^2(0))}{s(\l)}\0\\
T_0(x,\l)&=&\T_0(x,\l) \T_0^{-1}(0,\l)=
e^{-mx^+\E_+}e^{-mx^-\E_-}\label{vuoto}
\bj
and hence, the gauge transformation induced by the element $f_N$
transforms the vacuum transport matrix $T_0$ into $T_N$.
Note that the $x^{\pm}$ independent matrix $S$ commutes with the connection
(\ref{con}) corresponding to the vacuum solution $\Phi=0$. We can choose
this matrix in such a way that the element $f_N$ has no singularities on
$\hat{\C}_N^+$ except of simple poles at the ramification points.
The explicit expression for such $f_N$ is
\a
f_N&=&
\frac{e^{\frac{\var_N}{2}H+\B_N}}{2\sqrt{\prod_{p=1}^{2N}(\l-\mu_p)}}
\left (\begin{array}{cc}
\prod_{j=1}^N (\l+(-)^N\ep_j(x))&
\prod_{j=1}^N (\l+(-)^N\ep_j(x))\\
{}~&~\\
\frac{s_N(\l)-l_N(\l)}{\prod_{j=1}^N
(\l-(-)^N\ep_j(x))}&
\frac{s_N(\l)-l_N(\l)}{\prod_{j=1}^N
(\l-(-)^N\ep_j(x))}\end{array}\right )+\0\\
&~&\0\\
&+&\frac{e^{\frac{\var_N}{2}H-\B_N}}{2\sqrt{\prod_{p=1}^{2N}(\l+\mu_p)}}
\left (\begin{array}{cc}
\prod_{j=1}^N (\l-(-)^N\ep_j(x))&
-\prod_{j=1}^N (\l-(-)^N\ep_j(x))\\
{}~&~\0\\
-\frac{s_N(\l)+l_N(\l)}{\prod_{j=1}^N
(\l+(-)^N\ep_j(x))}&
\frac{s_N(\l)+l_N(\l)}{\prod_{j=1}^N
(\l+(-)^N\ep_j(x))}\end{array}\right )
\label{vest}
\b
where
\a
\B_N= m\l x^++\frac{mx^-}{\l}+\A_N
\label{def}
\b
Substituting the asymptotic expansion for $\A_N$ around the infinity and
the zero points into (\ref{vest})  and taking into account
(\ref{Jacobi}), (\ref{inf}) and (\ref{zero}) one gets the expansions
\ai
f_N(x,\l,s)&=&e^{\frac{\var_N}{2}H} \left( 1+X_{-1}+ O(\frac{1}{\l^2}) \right)
{}~~~~(\l,s)\rightarrow \infty_+ \label{gmeno}\\
f_N(x,\l,s)&=&e^{-\frac{\var_N}{2}H} \left( 1+ X_1+O(\l^2) \right)~~~~
(\l,s)\rightarrow \O_+ \label{gpiu}
\bj
from which together with (\ref{prod}) it follows that the expansion of
$f_N$  around the points $\infty_+$ and $\O_+$ can be identified with
the loop group elements $\tilde{g}_-$ and $\tilde{g}_+$  which
by gauge transformations generate the $N$--gap solution $\var_N$ from the
vacuum.
In \cite{Pa} we shall show that
in the soliton limit (\ref{vest}) reproduces  the
elements constructed in \cite{BaBe}.

\section{The dressing problem in the affine $\hat{SL}(2)$ group}

\setcounter{equation}{0}
\setcounter{footnote}{0}

In this section we shall demonstrate that there exists
a \dt\, $\hat{g}=\hat{g}_-^{-1}\hat{g}_+$
 which relates the vacuum solution
$(\var_0=0, \zeta_0=2m^2 x^+x^-)$ to the $N$--gap solution
$(\var_N,   \zeta_N)$ of (\ref{sinh})--(\ref{zeta}). We recall that
for  both solutions the field $\eta$ vanishes and that $\var_N$ is the
$N$--gap \shG\, solution (\ref{epphi}) considered in the previous section.
The field $\zeta_N$ which satisfies (\ref{zeta}) has still to be
determined. We outline our strategy. First, in view of the observation done
at the end of the previous section, the expansions of the element
(\ref{vest}) around the infinity point $\infty_+$ and the zero point $\O_+$
produce the loop group elements $\tilde{g}_-$ and $\tilde{g}_+$
which determine the \dt\, $\var_0 \rightarrow \var_N$. Further,
comparing (\ref{gmeno}) and (\ref{gpiu}) with (\ref{Xuno}) we shall get
a first order differential system for the field $\zeta_N$. Finally
we use (\ref{trucco})  to find $\hat{g}_{\pm}$. In order to guarantee the
consistency of this procedure we check that (\ref{inv}) is satisfied.

{}From (\ref{prod}) and (\ref{gmeno})--(\ref{gpiu}) it is seen that
in order to get $X_{-1}$  ($X_1$) one should  expand
$f_N$ up to the terms proportional to $\l^{-1}$ ($ \l$) when $(\l,s)
\rightarrow \infty_+$ ($(\l,s)\rightarrow \O_+$).
Inserting (\ref{Acal}) into (\ref{def}) and using the expansions
 (\ref{inf}), (\ref{zero}) as well as the equations
 (\ref{Jacobi}), (\ref{epphi})   we get
\a
\l \B_N&=& mx^-+\frac{mx^+}{2} \sum_{p=1}^{2N} \mu_p^2+
(-)^{N-1}\sum_{j=1}^N (\ep_j(x)-\ep_j(0)) +\0\\
&&+(-)^N \sum_{j=1}^N \int_{P_j(0)}^{P_j(x)} \frac{\mu^{2N} d\mu}{s_N(\mu)}
+O(\frac{1}{\l^2})~~~~~~~~~ (\l, s) \rightarrow \infty_+ \0\\
\l^{-1} \B_N&=& mx^+ +\frac{mx^-}{2} \sum_{p=1}^{2N} \frac{1}{\mu_p^2}+
(-)^N\sum_{j=1}^N \left( \frac{1}{\ep_j(0)}-\frac{1}{\ep_j(x)}\right)-\0\\
&&- \prod_{p=1}^{2N} \mu_p
\sum_{j=1}^N \int_{P_j(0)}^{P_j(x)} \frac{ d\mu}{\mu^2 s_N(\mu)}
 +O(\l^2) ~~~~~~~~~(\l,s)\rightarrow \O_+ \label{svbeta}
\b
On the other hand (\ref{ele}) is a polynomial of degree $2N-1$ on the
spectral parameter
\a
l_N(\l)=\l^{2N-1}\frac{\dpiu \var_N}{m}+ \ldots +(-)^{N+1} \l
\prod_{p=1}^{2N} \mu_p \frac{\dmeno \var_N  }{m}
\label{poli}
\b
Substituting the expansions (\ref{svbeta}) and (\ref{poli})
in (\ref{vest}) one obtains
\ai
X_{-1}&=& \left( mx^- +\frac{1}{2}\sum_{p=1}^{2N}(mx^+\mu_p+1)\mu_p+
(-)^N\sum_{j=1}^N \int_{P_J(0)}^{P_j(x)} \frac{\mu^{2N}
d\mu}{s_N(\mu)} +(-)^N \sum_{j=1}^N \ep_j(0) \right) \E_-+\0\\
&&-\frac{\dpiu \var_N}{m} \l^{-1} E^- \label{meno}\\
X_1&=& \left( mx^+ +\frac{1}{2}(\frac{mx^-}{\mu_p}+1)\frac{1}{\mu_p}-
\prod_{p=1}^{2N}\mu_p\sum_{j=1}^N\int_{P_j(0)}^{P_j(x)}
\frac{d \mu}{\mu^2 s_N(\mu)}+ \sum_{j=1}^N \frac{1}{\ep_j(0)}
\right)\E_+-\0\\
&&+\frac{\dmeno \var_N}{m} \l E^- \label{piu}
\bj
which compared with    (\ref{Xuno}) produce  the system
\ai
\dpiu \zeta_N&=& \dpiu \var_N -m \sum_{p=1}^{2N} (mx^+\mu_p+1)\mu_p-
2 m (-)^N \sum_{j=1}^N \int_{P_J(0)}^{P_j(x)} \frac{\mu^{2N}
d\mu}{s_N(\mu)}-\0\\
&&-2m (-)^N \sum_{j=1}^N \ep_j(0) \label{zetapiu}\\
\dmeno \zeta_N &=& -\dmeno \var_N
-m \sum_{p=1}^{2N}(\frac{mx^-}{\mu_p}+1)\frac{1}{\mu_p}+
2m \prod_{p=1}^{2N} \mu_p \sum_{j=1}^N\int_{P_j(0)}^{P_j(x)}
\frac{d \mu}{\mu^2 s_N(\mu)}-\0\\
&&-2m \sum_{j=1}^N \frac{1}{\ep_j(0)} \label{zetameno}
\bj
The integrand of the third term of (\ref{zetapiu}) is
a differential of a second kind on $\hat{C}_N$ whose unique singularities are
second--order poles at the two infinity points $\infty_{\pm}$;
the differential in the corresponding term of (\ref{zetameno}) is
again a differential of a second kind with second--order poles at the
zero  points $O_{\pm}$.
The $x^-$--derivative of (\ref{zetapiu}) and the $x^+$--derivative
of (\ref{zetameno}) can be calculated explicitly with the help
of (\ref{moto}). The result is
\ai
m^{-2} \dmeno \dpiu \zeta_N&=&
\prod_{p=1}^{2N} \mu_p^{-1}
\left( 2 \sum_{j=1}^N \ep_j^{2N}(x)
\prod_{l\neq j} \frac{\ep_l^2(x)}{\ep_j^2(x)-\ep_l^2(x)}-
\prod_{l=1}^N \ep_l^2(x)\right)+\0\\
&&+\frac{\prod_{p=1}^{2N} \mu_p}
{\prod_{j=1}^N \ep_j^2(x)} \label{zetamenopiu}\\
m^{-2} \dpiu \dmeno  \zeta_N &=&
\prod_{p=1}^{2N} \mu_p \left( 2\sum_{j=1}^N \frac{1}{\ep_j^2(x)}
\prod_{l \neq j} \frac{1}{\ep_l^2(x)-\ep_j^2(x)}-
\prod_{l=1}^N \frac{1}{\ep_l^2(x)} \right)+\0\\
&&+\frac{\prod_{j=1}^N \ep_j^2(x)}{\prod_{p=1}^{2N} \mu_p}\label{zetapiumeno}
\bj
Note that  under the change
\a
\ep_j \rightarrow \frac{1}{\ep_j},~~~~~~~~~
\mu_p\rightarrow \frac{1}{\mu_p} \label{cambio}
\b
the r. h. s. of (\ref{zetamenopiu}) transforms into the r. h. s. of
(\ref{zetapiumeno}). In order to show that (\ref{zetapiu}),
(\ref{zetameno}) is integrable, we set in the identity
\a
\prod_{l=1}^N \frac{1}{x_l}= \sum_{j=1}^N \frac{1}{x_j}
\prod_{l\neq j} \frac {1}{x_l-x_j}\label{furbo}
\b
$x_j=\ep_j^{\pm 2}$. Comparing (\ref{zetamenopiu}) and (\ref{zetapiumeno})
we get
$\dpiu \dmeno \zeta_N= \dmeno \dpiu \zeta_N = m^2 (
e^{2\var_N}+e^{-2\var_N})$ and hence $(\var_N,\zeta_N)$ is an \ags\,
of the $sl(2)$ CAT model. To demonstrate that this solution is in
the orbit of the vacuum one still has to check  the identities
(\ref{inv}). A direct calculation based on (\ref{moto}) and
(\ref{zetapiu}), (\ref{zetameno}) shows that (\ref{inv}) reduces to
\ai
&&\sum_{j=1}^N\left( \sum_{p=1}^{2N} \frac{1}{\ep_j^2-\mu_p^2}+
2\sum_{k\neq j} \frac{1}{\ep_k^2-\ep_j^2}\right) s_N^2(\ep_j)
\prod_{l\neq j} \frac{1}{(\ep_l^2-\ep_j^2)^2}=\0\\
&&=-\sum_{p=1}^{2N} \mu_p^2 +2(-)^{N+1}\sum_{j=1}^{N} \ep_j^{2N}
\prod_{l\neq j} \frac{1}{\ep_l^2-\ep_j^2} \label{sommapiu}\\
&&\frac{1}{\prod_{p=1}^{2N} \mu_p^2}\sum_{j=1}^N
\left( \frac{2}{\ep_j^2}-\sum_{p=1}^{2N} \frac{1}{\ep_j^2-\mu_p^2}-
2\sum_{k\neq j} \frac{1}{\ep_k^2-\ep_j^2} \right) s_N^2(\ep_j)
\prod_{l\neq j} \frac{\ep_l^4}{(\ep_l^2-\ep_j^2)^2}=\0\\
&&=-\sum_{p=1}^{2N} \frac{1}{\mu_p^2}-2\sum_{j=1}^N \frac{1}{\ep_j^2}
\prod_{l \neq j} \frac{\ep_l^2}{\ep_l^2-\ep_j^2}\label{sommameno}
\bj
The above
 two identities are equivalent since the substitution
(\ref{cambio}) exchanges them and therefore it is sufficient to check only
(\ref{sommapiu}) which can be done using the same approach as in the proof
of (\ref{iduno}). This concludes the demonstration of the existence a \dgt\,
which transforms the vacuum into the $N$--gap solution $(\var_N, \zeta_N)$.
The corresponding \dge\,  $\hat{g}=\hat{g}_-^{-1} \hat{g}_+$
is given by (\ref{trucco}) where $\tilde{g}_-$ and $\tilde{g}_+$ are
the expansions of  (\ref{vest}) around the infinity point
$\infty_+$ and the zero point $\O_+$ of $\hat{C}_N$ respectively.

{\bf Acknowledgements}

It is a pleasure to thank L. Bonora for many stimulating discussions and
C. P. Constantinidis for the careful reading of the
manuscript. The author is grateful to SISSA and INFN, sez. di Trieste for
the hospitality. Partial financial support from  ICTP is also gratefully
acknowledged.

\end{document}